\documentclass[reprint, amsmath,amssymb, aps,superscriptaddress,floatfix,prb]{revtex4-2} 
\usepackage[left=2cm,top=3cm,right=2cm,bottom=4cm,head=1cm,a4paper]{geometry}    

\usepackage[utf8]{inputenc}

\usepackage{graphicx}				

\usepackage{braket}
\numberwithin{equation}{section}
\usepackage{bm}
\usepackage{lipsum}
\usepackage{appendix}
\usepackage{orcidlink}
\usepackage{hyperref}
\usepackage{mathtools}
\usepackage{bbm}
\hypersetup{
    colorlinks,
    citecolor=black,
    filecolor=black,
    linkcolor=black,
    urlcolor=black
}

\usepackage{tikz}
\usetikzlibrary{decorations.markings}
\usetikzlibrary{arrows,positioning} 

\usepackage[noframe]{showframe}
\usepackage{framed}
 {\endMakeFramed}
\definecolor{shadecolor}{gray}{0.9}

\newcommand{\Tr}{\ensuremath{\textup{Tr}}}
\newcommand{\1}{\ensuremath{\mathbbm{1}}}

\renewcommand{\Im}{\ensuremath{\textup{Im}}}

\begin{document}

\title{Quantised Bulk Conductivity as a Local Chern Marker}
\date{\today}

\author{Peru d'Ornellas\orcidlink{0000-0002-2349-0044}}
\affiliation{\small Blackett Laboratory, Imperial College London, London SW7 2AZ, United Kingdom}
\author{Ryan Barnett\orcidlink{0000-0002-5122-2856}}
\affiliation{\small Department of Mathematics, Imperial College London, London SW7 2AZ, United Kingdom}
\author{Derek K. K. Lee}
\affiliation{\small Blackett Laboratory, Imperial College London, London SW7 2AZ, United Kingdom}

\begin{abstract}

A central property of Chern insulators is the robustness of the topological phase and edge states to impurities in the system. Despite this, Chern number cannot be straightforwardly calculated in the presence of disorder. Recently, work has been done to propose several local analogs of the Chern number, called local markers, that can be used to characterise disordered systems. However, it was unclear whether the proposed markers represented a physically-measurable property of the system. Here we propose a local marker starting from a physical argument, as a local cross-conductivity measured in the bulk of the system. We find the explicit form of the marker for a non-interacting system of electrons on the lattice and show that it corresponds to existing expressions for the Chern number. Examples are calculated for a variety of disordered and amorphous systems, showing that it is precisely quantised to the Chern number and robust against disorder.

\end{abstract}

\maketitle

\section{Introduction}
Since the initial discovery of quantised Hall conductance~\cite{klitzing_new_1980, laughlin_quantized_1981, thouless_quantized_1982} and throughout the subsequent decades spent exploring novel topological phases of matter~\cite{haldane_model_1988,hasan_topological_2010}, the Chern number has been one of the central tools for understanding condensed matter systems outside of the Landau symmetry-breaking framework~\cite{berry_quantal_1984}. In the study of non-interacting Chern insulators, different Hamiltonians can be classified according to the Chern number of their bands. Two Hamiltonians with differing Chern numbers cannot be smoothly deformed into one another without crossing a point where the system becomes conductive, and systems with nonzero Chern numbers will always have conducting edge modes at their boundaries~\cite{qi_general_2006}. The Chern number has also been extensively used outside of topological insulators, an example being the Kitaev spin-liquid model where it indicates whether the system has Abelian or non-Abelian anyons~\cite{kitaev_anyons_2006}.\par
A defining characteristic of quantum Hall physics is the robustness of the edge states and Hall conductance to impurities in the sample. Despite this, the Chern number -- the central quantity in the TKNN invariant determining the Hall conductance~\cite{thouless_quantized_1982} -- cannot be calculated in a disordered material. The Chern number is calculated for a single band, and depends on the structure of the Brillouin zone. If a material lacks translational symmetry, it is not possible to apply Bloch's theorem, crystal momentum is not a good quantum number and we have no concept of momentum space. Disorder-resistant methods exist~\cite{niu_quantized_1985, bellissard_noncommutative_1994}, however they compute a global Chern number, and cannot capture the local properties of materials with compound structure. This tension, that our tools for describing disorder-resistant physics are themselves undermined by disorder, suggests that our mathematical framework is incomplete. \par 
One possible solution to this problem is the development of local Chern markers. These attempt to translate the Chern number into a mathematical language that is resolved in real space, ensuring it is well-defined when the system has no translational symmetry. Several candidates for local markers exist, such as the Chern marker~\cite{bianco_mapping_2011}, the Bott index~\cite{loring_guide_2019}, and the Chern number defined by Kitaev in appendix C of~\cite{kitaev_anyons_2006}. Each of these markers has been derived by finding ways of re-expressing the Chern number in terms of quantities that can be evaluated in real space. This ensures that in an ideal uniform material with no boundaries -- where the Chern number can be calculated directly -- the marker should exactly equal the Chern number. Despite satisfying this requirement, existing markers suffer from two main drawbacks. Firstly, they display unexplained behaviour when taken out of the context where the Chern number is already well-defined. The Chern marker and Bott index have sharp drops around the boundary of any topological region~\cite{caio_topological_2019}, and Kitaev's local Chern number vanishes in non-infinite systems. Secondly, all markers were developed from a mathematical restatement of the Chern number, and the resulting expression does not obviously correspond to any physical quantity. Thus, it is not clear at all how one might attempt to measure it in a real system, or even if there is a way to connect them to an observable at all -- although some attempts have been made~\cite{bianco_orbital_2013, mitchell_amorphous_2018, rauch_geometric_2018}. \par
We present a derivation for a local marker that starts from physical grounds, defined as a localised version of the Hall conductivity~\cite{kubo_statistical-mechanical_1957}. Local current is measured around the position $\textbf R$ of a crosshair placed in the bulk of the material, motivating the name `crosshair marker'. The derived quantity is close to the local Chern marker defined by Kitaev \cite{kitaev_anyons_2006}, given by
\begin{align}
    C(\textbf R) = 4\pi \Im \Tr_{\textup{Bulk}} 
    \left ( 
    P\vartheta_{R_x} P \vartheta_{R_y} P
    \right ),
\end{align}
where $P$ is a projector onto the occupied band and $\vartheta$ is a step function at $\textbf R$ in the $x$ or $y$ directions. The trace is over a region around $\textbf R$ in the bulk. Since the marker has a straightforward interpretation, its behaviour around edges of the system can be understood intuitively in terms of current induced in the material. Furthermore, when evaluated in a disordered system, the crosshair marker is almost exactly quantised provided that it is measured sufficiently far from any edge modes. In contrast, the Bott index and Chern marker are not strictly quantised in the presence of disorder. We also find a connection between the crosshair marker and the Chern marker. Summing over all possible positions of the crosshair itself, we exactly recover the Chern marker, allowing for the physical interpretation of the crosshair marker to be extended to the Chern marker.\par
The paper is structured as follows: In \textsection \ref{sec:preliminaries}, we derive the necessary pre-requisites to understand the marker. These are threefold, in \textsection \ref{sec:fields} we describe the formalism for modelling the effect of electric fields on our system, in \textsection \ref{sec:current} we define a set of current operators on a lattice, and in \textsection \ref{sec:kato} we describe Kato's formalism for adiabatic quantum evolution based on~\cite{kato_adiabatic_1950}. In \textsection \ref{sec:marker_derivation} we use these concepts to derive the crosshair marker and discuss its connection to the Chern marker. Finally, in \textsection \ref{sec:examples} we give examples of the crosshair marker in the Qi-Wu-Zhang (QWZ) model with spatially-varying parameters, as well as an extension of the QWZ Hamiltonian to amorphous lattices. These serve as a testing ground for extending the Chern number to systems that are inaccessible to the conventional momentum space calculations.\par

\section{Preliminaries} \label{sec:preliminaries}
We work with a general non-interacting Hamiltonian on a two-dimensional tight-binding lattice. The system consists of set of $N$ sites, with positions $\textbf r_i$ arranged on either a regular or amorphous lattice. Each site has an internal degree of freedom hosting $\eta$ states. The Hamiltonian can be written in the form
\begin{align} \label{eqn:hamiltonian}
    H = \sum_{i,j}  \ket{\textbf {r}_i}\bra{\textbf r_j} \otimes H_{i,j},
\end{align}
where $H_{i,j} = H_{j,i}^\dag$ acts on the internal degrees of freedom. Here we will only work with systems in open boundary conditions. \par

\subsection{Electric Fields} \label{sec:fields}

\begin{figure}
    \centering
    \includegraphics[width = 0.7\columnwidth]{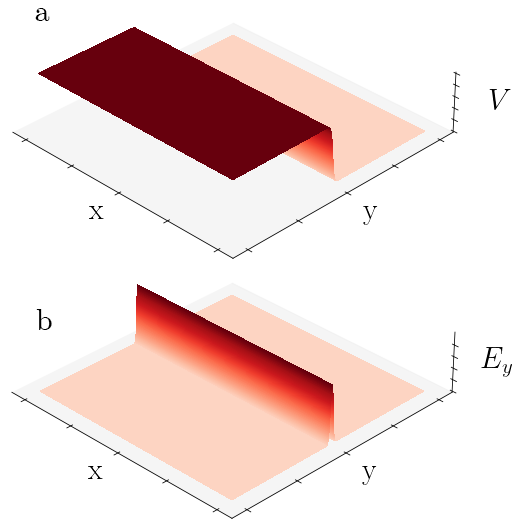}
    \caption{\label{fig:fields} a) The electric potential used to generate our electric field. b) The electric field, with only the nonzero $y$-component is shown.
    }
\end{figure}

In \textsection \ref{sec:marker_derivation} we examine the current generated by step-function electric potentials. Thus, here we precisely define the form of the electric field and potential, and study how they are represented on the lattice.\par
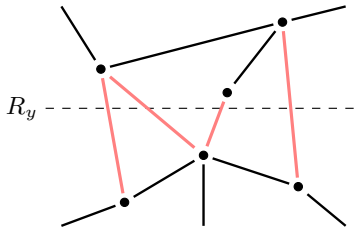
\begin{figure} 
    \resizebox{0.6\columnwidth}{!}{\begin{tikzpicture}

\def \squaresize{2}
\def \squarepos{0}
\def \arrowdistance {0.3}
\def \stepnum {4}
\def \len  {0.2}


\draw [dashed]( - 2 , 0) --  ( 2 , - 0);

\node (a) at (1,1.1) {};
\filldraw (a) circle (1.5pt);
\node (b) at (0.3,.2) {};
\filldraw (b) circle (1.5pt);
\node (c) at (0.,-0.6) {};
\filldraw (c) circle (1.5pt);
\node (d) at (-1.3,0.5) {};
\filldraw (d) circle (1.5pt);
\node (e) at (-1,-1.2) {};
\filldraw (e) circle (1.5pt);
\node (f) at (1.2,-1) {};
\filldraw (f) circle (1.5pt);

\node at (-2.3,0) {$R_y$};

\draw[line width=0.3mm] (a) -- (b);
\draw[line width=0.3mm] (a) -- (d);

\draw[red!50,line width=0.4mm] (c) -- (b);
\draw[red!50,line width=0.4mm] (c) -- (d);
\draw[red!50,line width=0.4mm] (d) -- (e);
\draw[red!50,line width=0.4mm] (a) -- (f);

\draw[line width=0.3mm] (c) -- (f);
\draw[line width=0.3mm] (c) -- (e);

\draw[line width=0.3mm] (d) -- (-1.8,1.3);
\draw[line width=0.3mm] (a) -- (1.8,1.3);
\draw[line width=0.3mm] (e) -- (-1.8,-1.5);
\draw[line width=0.3mm] (f) -- (1.8,-1.5);
\draw[line width=0.3mm] (c) -- (0,-1.5);

\end{tikzpicture}}
    \caption{A section of amorphous lattice is shown. All hopping terms in the Hamiltonian that cross the line $y = R_y$ acquire a phase given by eqn.~\ref{eqn:phase}, these are depicted in red. Unaffected hopping terms are shown in black. Note that the phase is ill-defined when a site $\textbf r_i$ lies on the line.}\label{fig:phase_lines}
\end{figure}

We wish to model the effect of raising the potential of one half of the system with respect to the other, resulting in an electric field across the boundary separating the two regions. Without loss of generality, let us consider raising the potential of the region below a dividing line at position $y = R_y$ with respect to the region above the line. In the continuum, this corresponds to an electric potential of the form $V(\textbf r) = -V_0 \theta(y - R_y) $, where $\theta$ is a Heaviside step function. The electric field acts over the line at $y = R_y$, as shown in figure \ref{fig:fields},
\begin{align}
    \textbf E(\textbf r)= V_0 \delta(y -R_y) \textbf e_y,
\end{align}
where $\textbf e_y$ is a unit vector in the $y$ direction. We transform to a gauge with zero scalar potential, representing the electric field with a magnetic vector potential,
\begin{align}
    \textbf A(\textbf r, t) = -V_0 t \delta(y - R_y) \textbf e_y.
\end{align}
Making the assumption that $V_0$ is small, we shall use Peierls substitution to describe the effect of a slowly varying magnetic vector potential on our lattice system~\cite{peierls_zur_1933}. The translation operators are modified by a Peierls phase $\ket{\textbf r_j}\bra{\textbf r_i} \rightarrow \ket{\textbf r_j}\bra{\textbf r_i} e^{i\alpha(\textbf r_i, \textbf r_j)}$, given by 
\begin{align}
    \alpha(\textbf r_i, \textbf r_j) = \int_{\textbf r_j}^{\textbf r_i} \textbf A(\textbf r) \cdot d\textbf r.
\end{align}
Note that we work in natural units, setting the electronic charge $q = \hbar = 1$. Given the form of the magnetic vector potential, we can see that translation operators only pick up a phase if they cross the line $y = R_y$. This is illustrated in fig.~\ref{fig:phase_lines}. The Peierls phase can be expressed as
\begin{align}\label{eqn:phase}
    \alpha(\textbf r_i, \textbf r_j) = A(t) \left [ \theta( y_{i} - R_y) - \theta( y_{j} - R_y) \right ],
\end{align}
with $A(t) = V_0 t$. Thus, the Hamiltonian is modified according to
\begin{align}\label{eqn:H_shift}
    H(A) = e^{i A(t) \vartheta_{R_y}} H e^{-i A(t) \vartheta_{R_y}},
\end{align}
where we have defined the projector onto the half space above $y = R_y$ as
\begin{align}\label{eqn:heaviside_def}
    \vartheta_{R_y} = \sum_{i} \theta(y_i - R_y) \ket{\textbf r_i} \bra{\textbf r_i}.
\end{align}
In the limit of small $A$, this can be expanded to first order,
\begin{align}\label{eqn:H_A_first_order}
    H(A) = H - i A(t) \left [ H, \vartheta_{R_y} \right ].
\end{align}
As we shall see in the next section, the change in the Hamiltonian is expressed in terms of a current operator for flow across the line $y = R_y$.\par
Note that for a static $A(t)$ in open boundaries, the $A$-dependence can be completely removed by a gauge transformation, so there will be no physical effect on our system. This is reflected by the fact that $\textbf E$ depends on $\partial_t A$, so static $A$ corresponds to zero electric field.

\subsection{Current Operators}\label{sec:current}

We construct an operator representing the flow of particles across a line bisecting the system. Again, without loss of generality, we will look at the flow of particles in the $x$-direction, across the vertical boundary at $x = R_x$. The number operator for particles to the right of this line is given by $\vartheta_{R_x}$, defined analogously to eqn. \ref{eqn:heaviside_def}. The time-evolution for the number of particles in this region is $\partial_t \langle \vartheta_{R_x} \rangle = i \langle [H, \vartheta_{R_x}] \rangle$. Thus, we can identify the operator representing flow of particles across the boundary as
\begin{align}\label{eqn:non_resolved_current}
    J_{R_x} = i[H, \vartheta_{R_x}].
\end{align}
This allows us to re-express eqn.~\ref{eqn:H_A_first_order} in the familiar form $H(A) = H - A J_{R_y}$, where $A$ is defined in equations \ref{eqn:phase}.\par
The operator $J_{R_x}$ evaluates the overall flow across the whole system, however in the following discussion we wish to separate the bulk and edge contributions. Thus let us decompose this operator into a sum of terms local to some point at $\textbf r$, according to
\begin{align}\label{eqn:current_operator}
    J_{R_x}(\textbf r) = \frac{1}{2} \left (
    \delta_{\textbf r} J_{R_x} +
    J_{R_x} \delta_{\textbf r}  \right ),
\end{align}
with $\delta_{\textbf r} = \ket{\textbf r}\bra{\textbf r}$. The two terms evaluate the contributions to $J_{R_x}$ from hopping into site $\textbf r$, and from hopping out of site $\textbf r$ respectively. The factor of $1/2$ is included to ensure that $\sum_{\textbf r}J_{R_x}(\textbf r) = J_{R_x}$, since every pair of sites is counted twice. This decomposition is not unique --  there are many ways one could express $J_{R_x}$ as a sum over local terms. However, the final result will be insensitive to particular method since we shall sum over all the bulk contributions, which will be well-separated from edge contributions.

\subsection{Kato Dynamics}\label{sec:kato}
We will work in the adiabatic limit, with a Hamiltonian that changes over a long timescale $T$, where the local density of states in the bulk has an energy gap $\Delta$. We define the projector onto the occupied states at $t = 0$ as $P_0 = \sum_{\textup{occ}}\ket{\psi_i(0)} \bra{\psi_i(0)}$, where $\ket{\psi_i(0)}$ is an eigenstate of $H$ at $t = 0$. In general, time-evolution will result in a final state $P(t)$ that is related to the initial state by the Schr\"odinger equation. However. if the evolution is sufficiently slow (for $T\gg \Delta^{-1}$), we expect that $P_0$ will adiabatically evolve to the instantaneous projector, $P_{I}(t) =\sum_{\textup{band}}\ket{\psi_i(t)} \bra{\psi_i(t)} $, where $\ket{\psi_i(t)}$ are eigenstates of $H$ at time $t$~\cite{avron_adiabatic_1987}. To ensure that our time-evolution is exactly adiabatic, we use a formalism based on that introduced in~\cite{kato_adiabatic_1950}, although a modern description can be found in~\cite{simon_tosio_2018,avron_adiabatic_1987,avron_adiabatic_1999}. We sketch out the basics here. However for a detailed discussion, see appendix \ref{sec:kato_appendix}. \par 
Rather than using $H$ to generate time evolution, we define an adiabatic Hamiltonian, $K$, which satisfies the instantaneous Von-Neumann equation
\begin{align}\label{eqn:P_adiabatic_derivative}
    \partial_t P_I(t) = -i[K, P_I],
\end{align}
where the change in $P_I$ follows the instantaneous eigenstates of $H$. Using the identity $P \dot P P = 0$ (where dot denotes a time-derivative), we can verify that the equation is satisfied by $K$ of the form
\begin{align}\label{eqn:k_def}
    K(t) = i[\dot P_I , P_I].
\end{align}\par
The full time evolution of the projector follows the instantaneous eigenstates in the limit of large $T \gg \Delta$. Thus, in the following discussion, where we are working in the adiabatic regime, we may use eqn.~\ref{eqn:P_adiabatic_derivative} to describe the change in $P$, and drop the subscript on $P_I$, since all our time evolution is adiabatic. Having defined the adiabatic Hamiltonian, let us examine the exact form of $K$ for our applied electric field, as well as deriving an adiabatic equivalent of the current operator in eqn.~\ref{eqn:current_operator}.

\subsubsection{Adiabatic EM Fields}\label{sec:kato_field}

The time-dependence of our Hamiltonian is given by eqn.~\ref{eqn:H_shift}. We work in open boundaries, so the operator $e^{i A(t) \vartheta_{R_y}}$ behaves as a straightforward unitary rotation on the eigenstates. Note that in periodic boundaries, this operator would enforce twisted boundary conditions, shifting both the eigenstates and eigenenergies of $H$. Thus, the time-dependence of the projector is
\begin{align}
    P(t) = e^{i A(t) \vartheta_{R_y}} P_0 e^{-i A(t) \vartheta_{R_y}},
\end{align}
and the time-derivative is $\dot P =  i \dot A \left [ \vartheta_{R_y} , P \right ]$. This can be inserted into definition \ref{eqn:k_def} to arrive at the form of the adiabatic Hamiltonian,
\begin{align}\label{eqn:exact_k}
    K = \dot A \left ( P\vartheta_{R_y} Q +
    Q\vartheta_{R_y} P \right ),
\end{align}
where $Q = \1-P$ is the complement of the projector.

\subsubsection{Adiabatic Current}\label{sec:kato_current}

The adiabatic current operator is derived following an analogous argument to that in \textsection \ref{sec:current}. We find the change in particle number for the region $x>R_x$, given by $\vartheta_{R_x}$. In this case, however, we use the adiabatic Von-Neumann equation~\ref{eqn:P_adiabatic_derivative} to express the expectation value, $\partial_t \langle \vartheta_{R_x} \rangle = i \langle [K, \vartheta_{R_x}] \rangle$. Thus, we can identify an adiabatic current operator 
\begin{align}
    J_{R_x}^A = i [K,\vartheta_{R_x}].
\end{align}
As before, we can localise this operator to extract only the current hopping to and from a position $\textbf r$ according to 
\begin{align}\label{eqn:local_adiabatic_current}
    J^A_{R_x} (\textbf r) = \frac{1}{2} \{\delta_{\textbf r}, J_{R_x}^A \}.
\end{align}
In appendix \ref{sec:near_adiabatic} we discuss carefully how the non-adiabatic expression for current relates to the adiabatic form as we approach the adiabatic limit.

\section{Marker Derivation}\label{sec:marker_derivation}

Before we give the formal derivation, let us describe the physical intuition behind our topological marker, which can be understood as a localised version of the Hall conductivity. The standard Hall conductivity is measured by applying a uniform electric field to a 2D material and measuring the current induced perpendicular to the field. This current is quantised and corresponds to the Chern number of the Hamiltonian~\cite{laughlin_quantized_1981}. A disadvantage of this calculation is that it is necessarily global, with the electric field uniform over the system and the current measured across the whole sample. In a material with disorder or compound structure, the calculation gives no spatially-resolved information about its topological properties.\par
Our objective is to find an analogous quantity that can be spatially resolved, giving information about which parts of a complex material are responsible for topological behaviour. It is not obvious how one should localise the Hall conductivity: neither the electric field, nor the current operator can be easily localised to a point. Maxwell's laws prohibit one from creating an electric field at a single point only. Equally, on the lattice, it is difficult to write down a consistent expression for current at a point.\par
Despite that fact that neither electric field nor current can individually be made local to a point, it is possible to localise the overall Hall conductance. This is because both field and current can be localised to a line. By partitioning the system, and raising the electric potential of one part, we induce an electric field acting across the dividing line. Similarly, given a partition, by measuring the transfer of electrons from one part to the other, we are able to determine the current across the dividing line.\par 

\begin{figure} [t]
    \resizebox{0.75\columnwidth}{!}{\begin{tikzpicture}

\def \squaresize{2}
\def \squarepos{0}
\def \arrowdistance {0.4}
\def \stepnum {4}
\def \len  {0.2}

\filldraw[white,inner color=red!70,outer color=white] (0,0) circle (1.3);

\draw [dashed] ( - \squaresize , -\squaresize) --  ( \squaresize , -\squaresize) --   ( \squaresize , \squaresize) -- ( - \squaresize, \squaresize) -- ( - \squaresize , -\squaresize);

\draw (0,- \squaresize-0.1 ) -- (0, \squaresize+0.1 ) ;

\draw (- \squaresize-0.1 ,0) -- ( \squaresize+0.1, 0 ) ;

\foreach \x in {-4,...,4}
{
\draw [-latex,blue] (  \x * \squaresize / 4.3-0.1 , - \len)  to [out=90,in=200] ( \x * \squaresize / 4.3+0.3-0.1 , \len);

}

\draw [-latex, gray] (-0.3 , 1.15*\squaresize) -- (0.3 , 1.15*\squaresize) ;
\draw node [] at (0.0 , 1.3*\squaresize) {$J_{R_x}$};

\draw [-latex, gray] (  -1.15*\squaresize ,-0.3) -- ( -1.15*\squaresize, 0.3) ;
\draw node [] at (-1.3*\squaresize, 0) {$\bf E$};

\draw node [] at (0, -\squaresize-0.3) {$ R_x $};
\draw node [] at (\squaresize+0.4, 0) {$ R_y $};


\end{tikzpicture}}
    \caption{A schematic for local Hall measurement. The bottom half of the system has its potential raised with respect to the top half, generating an electric field acting over the horizontal line. Current is measured between the left and right halves, across the vertical line. The flow of charge is represented in blue, showing the curved path due to the presence of a magnetic field. The red shading shows the region around the crosshair where a nonzero contribution to the marker is measured.}\label{fig:crosshair_quarters}
\end{figure}
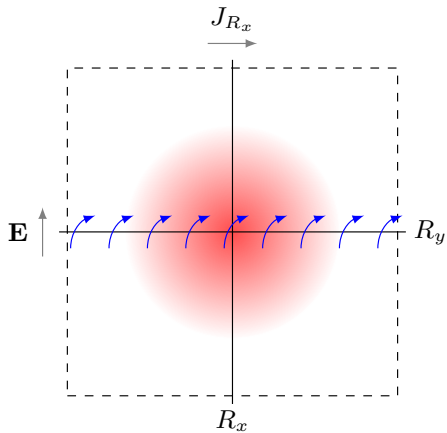

The route to a local Hall current is to partition the system along two perpendicular dividing lines, as shown in fig.~\ref{fig:crosshair_quarters}. An electric field is induced between the top and bottom parts of the system by uniformly raising the voltage of the lower part. This will induce a vertical current acting across the dividing line. When the material has nonzero Chern number, a Hall current will also be induced from left to right. If the system is insulating (i.e.~gapped) we expect that all current should be localised to the region around the horizontal line. Measurement of the current is then taken over the vertical line, catching only induced current in the horizontal direction - effectively only the Hall current. We expect all nonzero contributions to come from the region around the point where the lines cross - motivating the name `crosshair marker'.\par
As we shall see in the following examples, this argument fails when the system becomes conducting and the gap closes. In a conducting system, current is no longer locally constrained - an electron can easily tunnel across the entire system. Thus we expect that the marker will be nonzero even very far from the crosshair. Consequently, as well as the central peak, we expect to see an additional term wherever conducting edge states are present, at any distance from the crosshair. 

\subsection{Exact derivation}

We measure the local contribution to current across the line $x=R_x$ in the adiabatic regime. The current will be expressed in the form of a conductivity $J = \sigma E$, where $\sigma$ is to be determined. Our starting point is the expectation value of the current operator defined in \textsection\ref{sec:kato_current} for the case of a filled band of states,
\begin{align}
    \left \langle J^A_{R_x} (\textbf r) \right \rangle
    =\frac{1}{2}  \Tr \left [P  \{\delta_{\textbf r}, J_{R_x}^A \} \right ],
\end{align}
where $P$ projects onto a fully occupied band of states. By cycling terms in the trace and using the identity $\1 = P+Q$, we can split this expression into two terms
\begin{align} \label{eqn:j_val}
\begin{aligned}
    \left \langle J^A_{R_x}(\textbf r) \right \rangle & = \Tr_{\textbf r} 
    \left ( 
    P J_{R_x}^AP
    \right ) \\& +  
    \frac{1}{2}\Tr_{\textbf r} 
    \left ( 
    P J_{R_x}^A Q + QJ_{R_x}^AP
    \right ),
\end{aligned}
\end{align}
where $\Tr_{\textbf r} \vcentcolon= \Tr(\delta_{\textbf r}\ ... )$ denotes a local trace over the internal degrees of freedom at site $\textbf r$. We treat these two terms separately. The first term is a local marker of Chern number whereas the second term vanishes. Considering only the second term, let us insert the exact form of $J_{R_x}^A = -[[\dot P,P], \vartheta_{R_x} ]$ and use the Jacobi identity,
\begin{align} \label{eqn:cyclic}
\begin{aligned}
    \Tr _{\textbf r} \left ( P J_{R_x}^AQ  \right ) + h.c. &=
     \Tr _{\textbf r} 
    \left ( P  
    \left [[P, \vartheta_{R_x}],\dot P\right ]
    Q  \right ) \\ 
    &+
     \Tr _{\textbf r} 
    \left ( P  
    \left [[ \vartheta_{R_x},\dot P], P\right ]
    Q  \right )+ h.c.
\end{aligned}
\end{align}
Using the identities $P[P,\vartheta_{R_x}] =[P,\vartheta_{R_x}]Q $ and $P \dot P = \dot P Q$, we see that the first term here vanishes. We are left with the second term which can be written as
\begin{align}
    \Tr _{\textbf r} \left ( P J_{R_x}Q  \right ) + h.c. = 
    - \Tr _{\textbf r} 
    \left ( P  
    [ \vartheta_{R_x},\dot P]
    Q  \right )+ h.c.
\end{align}
Using $Q = \1-P$, we may re-express this in the form,
\begin{align}
    \Tr _{\textbf r} \left ( P J_{R_x}Q  \right ) + h.c. = 
    - \Tr _{\textbf r} 
    \left ( P  
    [ \vartheta_{R_x},\dot P]
    \right ) 
    + h.c.
\end{align}
Now, let us insert the exact form for $\dot P$ for a system undergoing adiabatic application of the step-function electric field defined in \textsection\ref{sec:kato_field}: $\dot P = i\dot A [\vartheta_{R_y} , P]$. We get a final expression
\begin{align}\label{eqn:constituent_vanish}
\begin{aligned}
    \Tr _{\textbf r} \left ( P J_{R_x}Q  \right ) &+ h.c.\\ &=- i \dot A \Tr _{\textbf r} 
    \left ( P  
    \left [ \vartheta_{R_x},[\vartheta_{R_y}, P]\right ]
    \right ) + h.c.
\end{aligned}
\end{align}
Writing out the terms in the trace explicitly, we get
\begin{align}
\begin{aligned}
    i\Tr _{\textbf r} 
    \left ( P  
    \left [ \vartheta_{R_x},[\vartheta_{R_y}, P]\right ]
    \right ) &= i\Tr_{\textbf{r}} (P\vartheta_{R_x}\vartheta_{R_y}P )\\
    &- i\Tr_{\textbf{r}} (P\vartheta_{R_x}P\vartheta_{R_y})\\
    &- i\Tr_{\textbf{r}} (\vartheta_{R_x}P\vartheta_{R_y}P)\\
    &+ i\Tr_{\textbf{r}} (P\vartheta_{R_x}\vartheta_{R_y}).
\end{aligned}
\end{align}
The step functions commute with one another, and with $\delta_{\textbf r}$. This means that every constituent term in eqn.~\ref{eqn:constituent_vanish} is anti-Hermitian due to the factor of $i$. Thus the whole expression vanishes when the Hermitian conjugate is added, 
\begin{align}
    \Tr _{\textbf r} \left ( P J_{R_x}Q  \right ) &+ h.c. = 0. 
\end{align}
Now, let us return to eqn \ref{eqn:j_val}. We have shown that the second term vanishes, so we are left with
\begin{align} 
    \left \langle J_{R_x}(\textbf r) \right \rangle & = \Tr_{\textbf r} 
    \left ( 
    P J_{R_x}P
    \right )\\
    &=i \Tr_{\textbf r} 
    \left ( 
    P [K, \vartheta_{R_x}] P
    \right ).
\end{align}
We use expression \ref{eqn:exact_k} to insert the exact form of $K$
\begin{align} 
    \left \langle J_{R_x}(\textbf r) \right \rangle & = i \dot A \Tr_{\textbf r} 
    \left ( 
    P\vartheta_{R_x} Q \vartheta_{R_y} P
    \right ) + h.c.
\end{align}
Using the fact that $\dot A = -E$, we see that this expectation value has taken the form of a conductance, $\langle J \rangle = \sigma E$. We also use $Q = \1 - P$ to get the conductance in the form
\begin{align}\label{eqn:marker_form}
    \sigma(\textbf r; \textbf R) = 2 \Im \Tr_{\textbf r} 
    \left ( 
    P\vartheta_{R_x} P \vartheta_{R_y} P
    \right ).
\end{align}\par
Finally we insert a factor of $2\pi$ to get the final expression for the marker
\begin{align}
    C(\textbf r;  \textbf R) = 4\pi \Im \Tr_{\textbf r} 
    \left ( 
    P\vartheta_{R_x} P \vartheta_{R_y} P
    \right ).
\end{align}
This quantity is analogous to the Chern marker presented in appendix C of~\cite{kitaev_anyons_2006}, although where Kitaev used a global trace, here we have a local trace. Kitaev's marker was shown to exactly equal the Chern number in an infinite system. For finite system size, however, his marker vanishes due to edge-state contributions. By replacing the global trace with a local trace, we are here able to separate the bulk contribution from the edge contribution, allowing for the marker to be calculated in a finite system. We can extract the quantised Chern number by only summing over the bulk contributions -- found at $\textbf r$ close to $\textbf R$ -- and ignoring the edge-state terms at $\textbf r$ far from $\textbf R$.\par
We note two important caveats. Firstly the marker is only strictly well-defined as the sum of the bulk contributions around $\textbf R$, rather than as a function of $\textbf r$. This is because there is a degree of arbitrariness to how the current was made local to a point in definition \ref{eqn:current_operator}, since the purpose is to separate the bulk and edge contributions before summing only the bulk terms. Physically, the summed marker corresponds to the total current measured in the bulk over the line at $x = R_x$. \par 
 The second caveat is that the adiabatic current defined in eqn.~\ref{eqn:local_adiabatic_current} only captures the current generated by the change in $P$ over time. However, it is also possible for a system to have persistent currents, i.e. flow between occupied states that does not change the projector itself. Since this type of circulating current cannot change the number density at any site, it does not contribute to the net flow of electrons into a given region. Despite this, it can still contribute terms to the current defined in eqn.~\ref{eqn:current_operator}. Thus we must also require that the system has no persistent currents at any point in the adiabatic time evolution. This effect is explained in detail in appendix \ref{sec:near_adiabatic}.\par
Finally it is worth mentioning that the crosshair marker can be used to calculate the Chern marker, $\mathfrak{C}$, given in~\cite{bianco_mapping_2011} according to
\begin{align}
    \mathfrak{C}(\textbf r) = \int d^2 \textbf R\ C(\textbf r;  \textbf R),
\end{align}
where the integral is over the whole system. The proof of this is given in appendix \ref{sec:chern_marker_connection}.

\section{Examples}\label{sec:examples}

Examples are calculated for two sets of topological quantum systems. The first will be the Qi-Wu-Zhang (QWZ) model on a square lattice, the simplest example of a Chern insulator with only nearest neighbour interactions~\cite{qi_topological_2006,asboth_short_2016}. The second example will be an extension of the QWZ model to amorphous lattices.

\begin{figure*}
\centering
\includegraphics[width = 0.8\paperwidth]{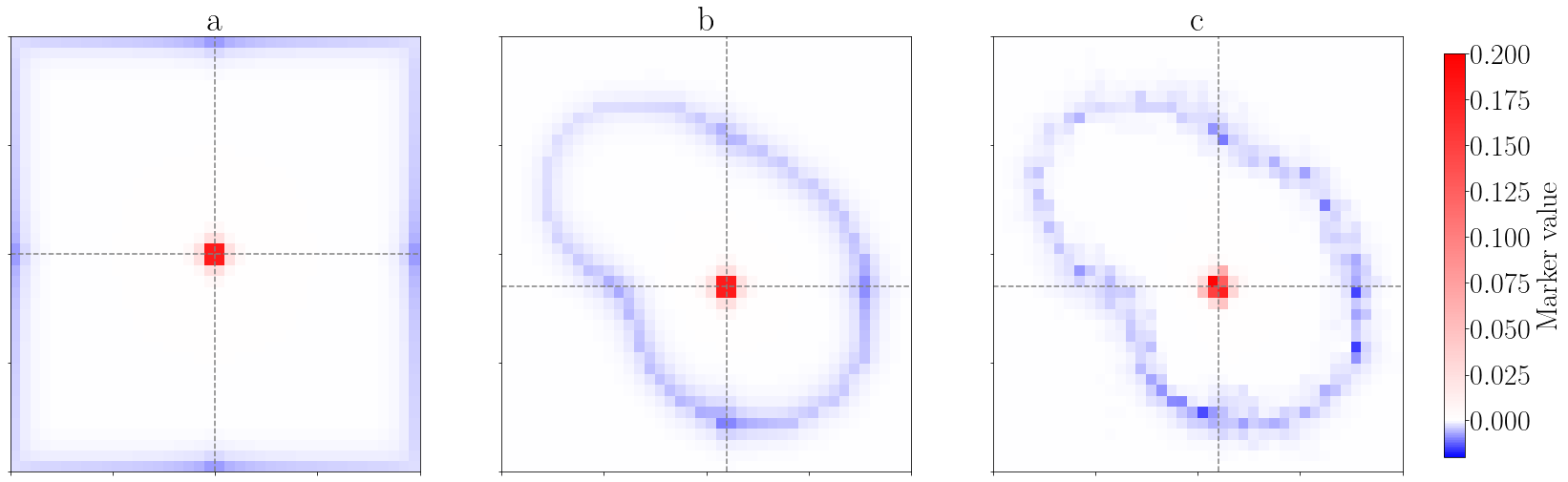}
\caption{\label{fig:three_plots}
Crosshair marker, C(\textbf r;  \textbf R), plotted as a function of $\textbf r$ for three example systems on 40x40 lattices. Position of the crosshairs (determined by $\textbf R$) is indicated with dashed lines, where $\textbf R$ indicates the point where the lines meet. a) A uniform QWZ system with $u = 1.6$. b) A compound system with $u = 1.6$ inside the central region and $u = 2.6$ outside. The boundary of the region follows the blue ring, which indicates the presence of conducting edge states. c) A compound system, as in (b), however $u$ has an added per-site Gaussian noise term with standard deviation $\sigma = 0.5$.}
\end{figure*}

\subsection{Qi-Wu-Zhang Model}\label{sec:QWZ_model}

We work on a square lattice of size $(L_x, L_y)$, where site positions are given by $\textbf r_i = (n_x,n_y) \in \mathbb{Z}^2$. Each site has two internal degrees of freedom. The Hamiltonian is
\begin{align}\label{eqn:qwz_hamiltonian}
    H &= \sum_{\textbf r} \ket{\textbf r+ 1_x}\bra{\textbf r} \otimes \frac{\sigma_z + i\sigma_x}{2} + h.c. \nonumber \\
    &+ \sum_{\textbf r} \ket{\textbf r+ 1_y}\bra{\textbf r} \otimes \frac{\sigma_z + i\sigma_y}{2} + h.c. \nonumber \\
    &+ \sum_{\textbf r} \ket{\textbf r}\bra{\textbf r} \otimes u \; \sigma_z,
\end{align}
where $\sigma_i$ are the Pauli matrices. The parameter $u$ determines the topological properties of the Hamiltonian. In a uniform system, the Chern number is $-1$ for $-2 < u < 0$, $+1$ for $0 < u < 2$ and zero otherwise. In the following examples we break translational symmetry by allowing $u$ to vary on a per-site basis, $u \rightarrow u_{\textbf r}$. \par
We calculate the crosshair marker for three example systems, shown in figure \ref{fig:three_plots}. The first system is a uniform lattice with open boundaries. In the second we show a compound system with open boundaries and a central region that we expect to have Chern number $+1$, surrounded by a region with Chern number 0. In all cases, the crosshair is placed in the bulk of the topological region and the expected Chern number is +1. Close to the crosshairs, there is a positive contribution from the local conductance as derived in \textsection\ref{sec:marker_derivation}. Around the boundary of the topological region there is a negative contribution due to the presence of edge states. The overall trace of the marker vanishes, so this negative contribution sums to the same magnitude as the central peak.\par
\begin{figure}
\centering
\includegraphics[width = 0.8\columnwidth]{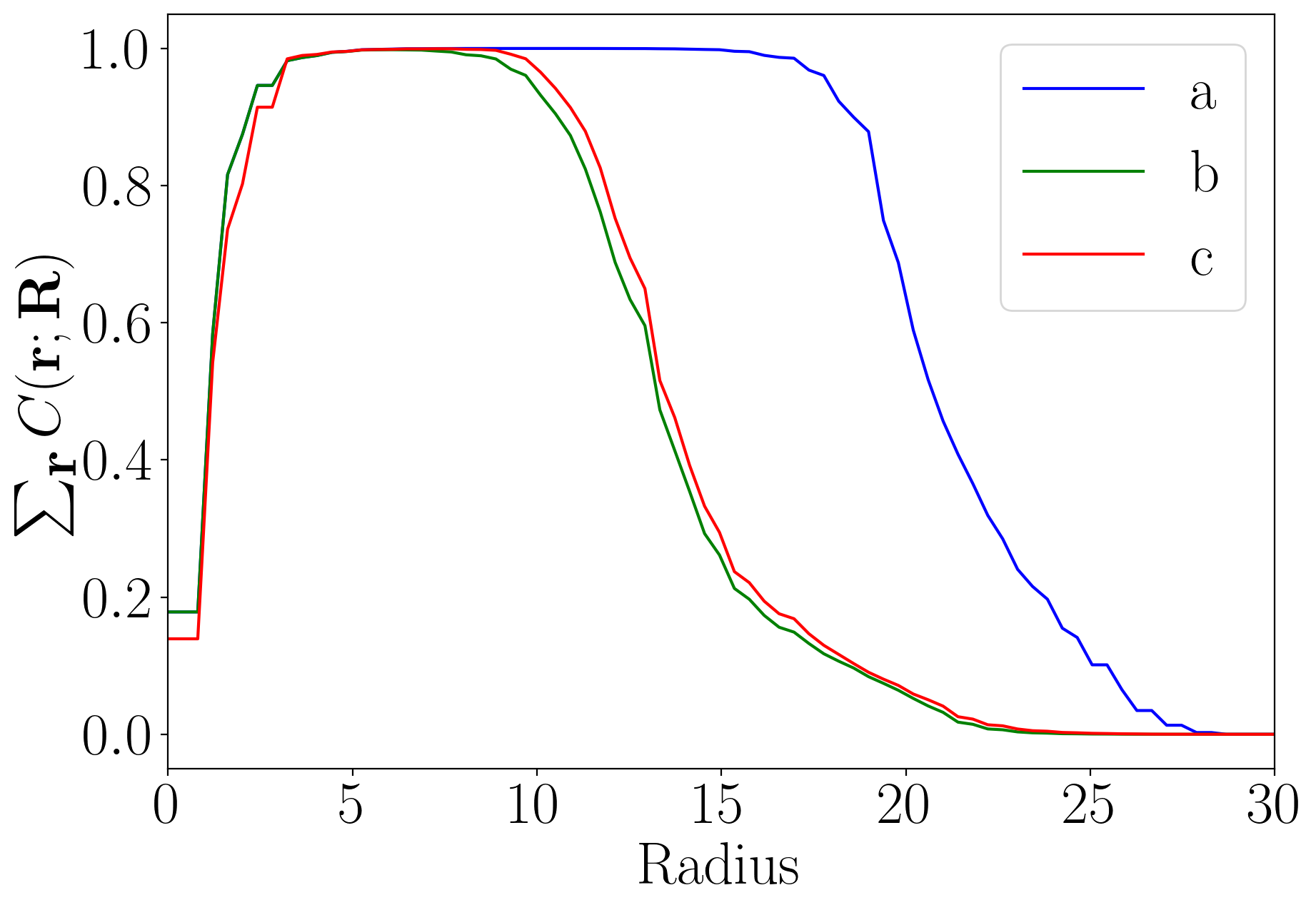}
\caption{\label{fig:qwz_sums}
Plots of the sum of the crosshair marker in $\textbf r$ for a circle of a given radius around the crosshair position $\textbf p$. Plots are shown for the three examples in fig. \ref{fig:three_plots}. The maximum values reached in each case are 0.99997, 0.9981 and 0.9989 respectively.}
\end{figure}

To demonstrate the quantisation of the marker, we sum the value enclosed within a radius around the central peak. When the radius is sufficiently large that it captures the whole peak the sum equals to the Chern number. If the radius includes contributions from the conducting edge states, the value will be reduced. This is demonstrated in fig. \ref{fig:qwz_sums}, where we have plotted the sum against the radius of the region being summed over. To estimate the quantised sum of the marker, we take the maximum value in this range, and see that this estimate is quantised to around one part in $10^3$.

\subsection{Amorphous Qi-Wu-Zhang Model}\label{sec:amorphous}

\begin{figure}
\centering
\includegraphics[width =0.95\columnwidth]{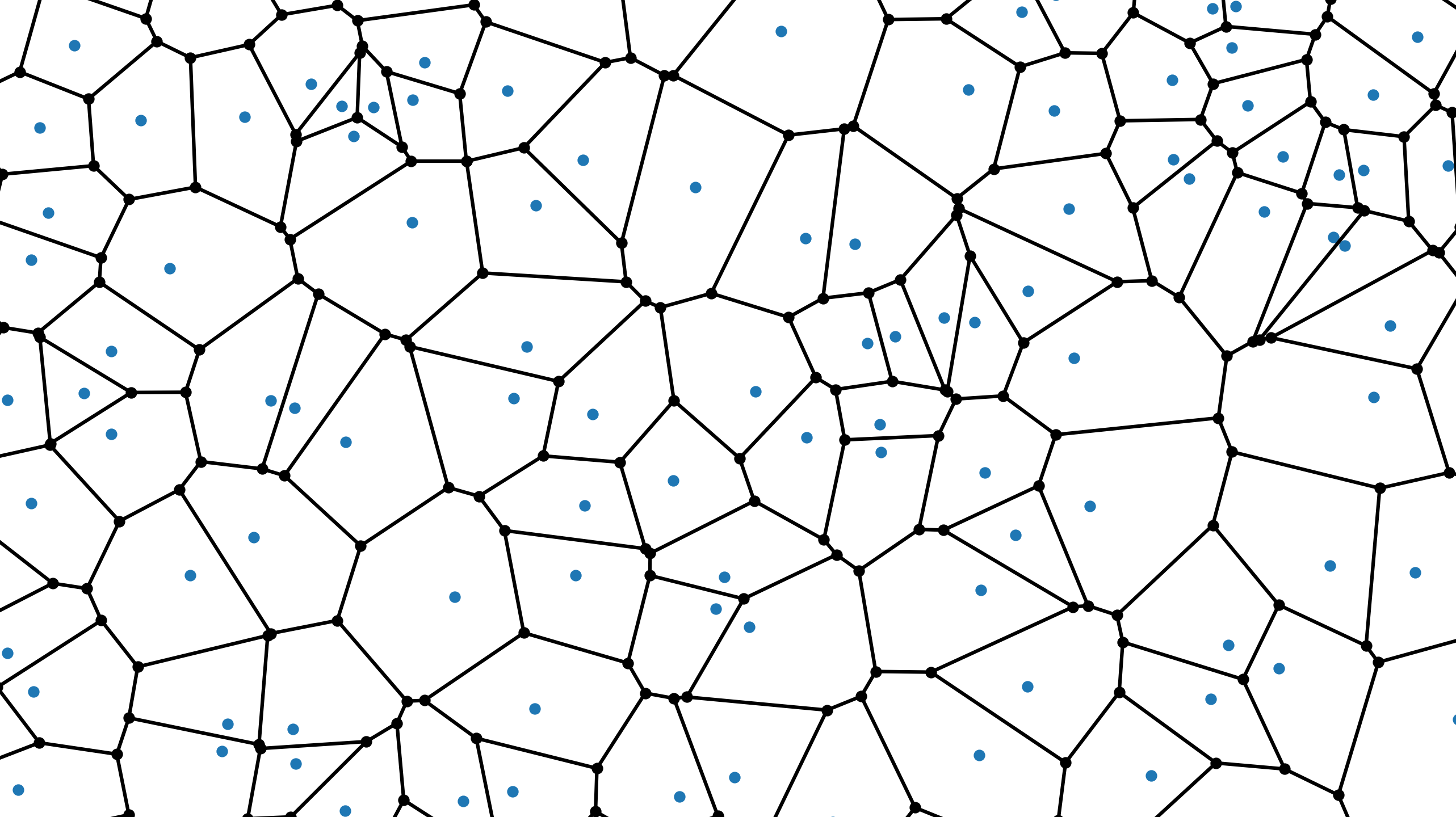}
\caption{\label{fig:lattice}
    An example section of amorphous lattice. Shown in blue are a set of points that have been randomly generated. In black we see the resulting Voronoi lattice, where each cell encloses the region of space that is closer to a given blue point than any other. Our sites (shown in black) are placed at the corners where three cells meet and connected with an edge along the boundary between two cells. }
\end{figure}

\begin{figure*}
\centering
\includegraphics[width = \textwidth]{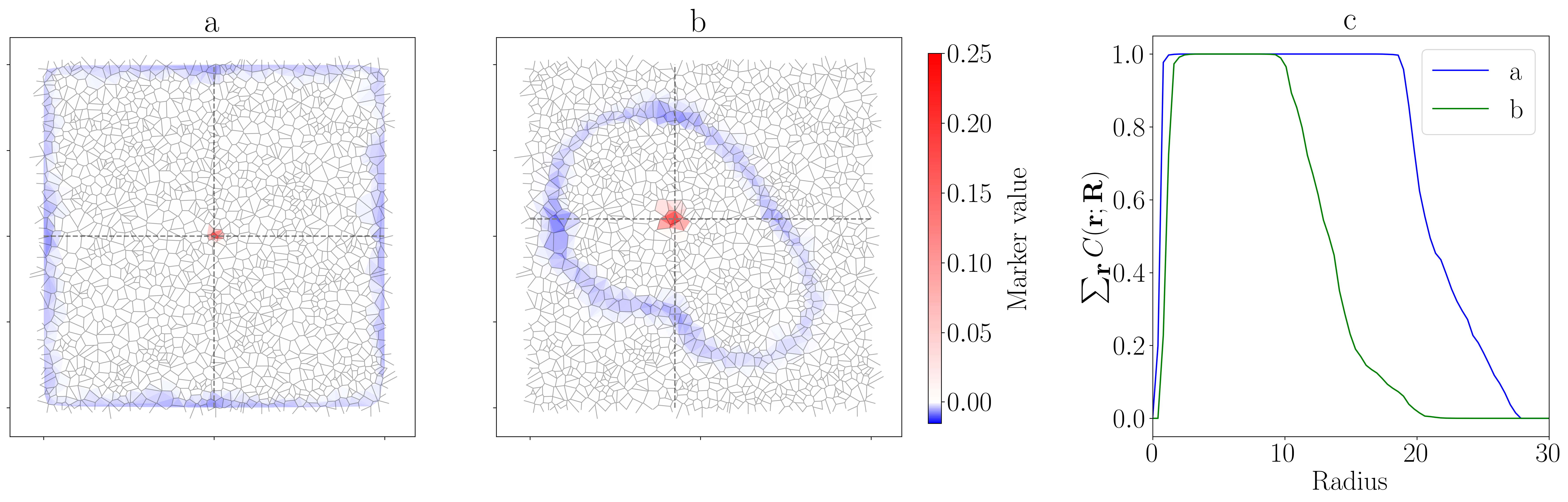}
\caption{ \label{fig:amorphous_plots}
Crosshair marker, C(\textbf r;  \textbf R), plotted as a function of $\textbf r$ for two example amorphous systems. As in fig.~\ref{fig:three_plots}, the position of the crosshairs (determined by $\textbf R$) is indicated with dashed lines. Both cases are for the same lattice with 3200 sites, decorated with a different Hamiltonian. a) The crosshair marker for the system with uniform $u = -1$. b) Crosshair marker for a compound amorphous system with $u  =-1$ inside the central region and $u  = -2$ outside. c) Plots of the sum of the crosshair markers for the two given systems. Maximum values reached in are 1.0007, and 1.00002 respectively }
\end{figure*}

We now look at extending the QWZ model beyond a grid to arbitrary amorphous lattices, providing an example of a system that is completely inaccessible to traditional methods for calculating Chern number. We prepare a set of arbitrary lattices using a Voronoi construction, identical to the construction of a Wigner-Seitz cell \cite{wigner_seitz_1933,florescu_designer_2009}. Starting with a set of random points, we may generate a cell for each point containing the region of space closer to that point than any other. A lattice is created by taking the barrier between Voronoi cells as the edges, and the vertices as the points where three adjacent Voronoi cells meet. This generates a lattice where every point is trivalent. An example lattice is shown in figure \ref{fig:lattice}.\par
Once a lattice has been chosen, the Hamiltonian is constructed using a smooth continuation of the QWZ model to account for edges that can point in an arbitrary direction. The Hamiltonian is given by
\begin{align}\label{eqn:amorphous_hamiltonian}
\begin{aligned}
    H &= \sum_{i,j} \ket{\textbf r_i}\bra{\textbf r_j} \otimes h_{\theta_{i,j}} + h.c. \\
    &+ \sum_{i} \ket{\textbf r_i}\bra{\textbf r_i} \otimes u \; \sigma_z.
\end{aligned}
\end{align}
Here, rather than the standard $x$ and $y$-hopping matrices from eqn. \ref{eqn:qwz_hamiltonian}, we use a matrix $h_\theta$ that depends on the angle $\theta$ that the edge $\textbf r_j \rightarrow \textbf r_i$ makes with the $x$-axis. This is defined as
\begin{align}
    h_\theta &= \frac{1}{2}\begin{pmatrix}
1 & ie^{-i\theta}\\ 
 ie^{i\theta}& -1
\end{pmatrix}.
\end{align}
Inserting into this expression $\theta = 0$ or $\theta = \frac{\pi}{2}$, we recover the standard QWZ hopping matrices. Also note that $h_{\theta + \pi} = h_{\theta }^*$.\par
As before, the parameter $u$ determines whether the system is in a topological or trivial state, although the phase does not have the same dependence on $u$ as before. By considering the band structure as a function of $u$, we determine numerically that the system is gapped and topological for $-3 < u < -1$. A justification for this is presented in appendix \ref{sec:amorphous_appendix}. \par
In fig. \ref{fig:amorphous_plots} we show crosshair markers for two example systems, as well as a plot of sums of the markers against the radius of the region summed over. As before, the bulk sum is quantised to around one part in $10^3$.

\section{Conclusion and Outlook}

We have developed a physically motivated local marker for the Chern number of a two-dimensional quantum system. The marker was derived as a local Hall conductivity for current around a crosshair in the bulk. We present a rigorous expression for this conductance for non-interacting electrons on the lattice, and show that it is equivalent to the real-space Chern number presented in~\cite{kitaev_anyons_2006}, although we make a slight modification to ensure it is nonzero for finite-size systems. Mathematically, the marker plays a similar role to existing methods such as Chern marker~\cite{bianco_mapping_2011} and Bott index~\cite{loring_guide_2019} -- locally indicating the topological phase of the system -- however unlike these markers, the crosshair marker is precisely quantised in the bulk.\par
The arguments used to derive the crosshair marker are essentially a prescription for measuring the quantity in the lab: a step-function potential is used to excite current across the system in the vertical direction. The cross-conductance is then measured in the horizontal direction. An obvious next step will be to look at experimentally verifying the proposed method. \par
Until recently, local markers were developed in a non-interacting context as a property derived from the projector onto a band of states. In interacting systems, where the ground state does not correspond to a projector onto a set of single-particle states, most existing markers are inapplicable, however early progress has been made in understanding how local topology can be detected in interacting systems~\cite{markov_local_2021}. Clearly, in such a system the mathematical formalism presented here does not work, however our physical argument does not depend on the microscopic details of the system. Thus, it would be interesting to re-express the marker in a many-body framework. Similarly, we expect that the physical definition of the marker could provide a clue to understanding the role that local topology has to play in systems out of equilibrium~\cite{mcginley_classification_2019,caio_topological_2019, golovanova_truly_2021}.

\section{Acknowledgments}
The author would like to thank Anton Markov, Charles Stahl, Gino Cassella, Joe Sykes and Tom Hodson for useful discussions. We acknowledge financial support from the UKRI Doctoral Training Partnership, with grant number EP/R513052/1.

\bibliography{references.bib}

\appendix
\section{Derivation of Adiabatic Time-Evolution Operator}\label{sec:kato_appendix}

In this section we will justify the form of the adiabatic Hamiltonian,
\begin{align}
    K(t) = i\left[ \partial_t P_I(t), P_I(t) \right],
\end{align}
as the generator of the correct adiabatic time-evolution for a given initial projector, $P_0$. We introduce a parameter $T$ that defines the period over which the system evolves, $t \in [0,T]$. Changing variables to $s = tT^{-1}$, we arrive at the scaled Schr\"odinger equation
\begin{align}\label{eqn:scaled_schrodinger}
    i \partial_s U_T(s) = TH(s) U_T(s),
\end{align}
where we can use $T$ to parametrise how slowly the system changes. We will also define the scaled adiabatic Hamiltonian
\begin{align}
    \bar K(s) = i[\partial_s P_I(s), P_I(s)].
\end{align}
Since $\partial_s = T \partial_t$, so this quantity is related to $K$ by $\bar K = T K$.\par  
Our aim will be to show that in the limit of large $T$, the time-evolved state $P(s) = U_T(s) P_0 U^{\dag}_T(s)$, can be approximated by the instantaneous projector, $P_I(s) = \sum_{i \in \textup{band}} \ket{\psi_i(s)} \bra{\psi_i(s)}$, where $\ket{\psi_i(s)}$ are eigenstates of $H(s)$. We wish to define an adiabatic time-evolution operator $U_A(s)$ such that
\begin{align}\label{eqn:U_A_def}
     P_I(s) = U_A(s) P_0 U_A^\dag(s).
\end{align}
We propose an ansatz for $U_A$ that satisfies the following conditions,
\begin{align}
    U_A(0) &= \1, \\
    i \partial_s U_A(s) &= \left [ TH(s) + \bar K(s) \right ] U_A(s).
\end{align}
To verify that $ U_A$ produces the right dynamics, let us consider the quantity
\begin{align}
    G(s) =  U_A(s)^\dag P_I(s) U_A(s).
\end{align}
In the interest of keeping equations legible, we will drop explicit $s$-dependence from our operators. From hereon in it is safe to assume that, unless stated otherwise, all operators are being evaluated at time $s$. We take the $s$-derivative of $ G$,
\begin{align}
    \partial_s  G &= \partial_s \left [ ( U_A^\dag  P_I)( P_I  U_A) \right ].
\end{align}
Using the identity
\begin{align}
    \partial_s ( P_I  U_A) = \partial_s P_I  U_A -i  P_I \left ( T H(s) + \bar K(s)  \right )  U_A,
\end{align}
we can show that
\begin{align}
    \partial_s  G &= 
    U_A^\dag \left ( iT [H,P_I] + \{P_I,\partial_s P_I\}  + i [\bar K,P_I] \right )U_A.
\end{align}
The first term in this expression vanishes, since $P$ and $H$ commute by definition. Next we can use the identities $\partial_s P_I = \{P_I, \partial_s P_I\}$ and $P_I \partial_s P_I P_I = 0$ to show that the second and third terms respectively equal $\partial_s P_I$ and $-\partial_s P_I$, cancelling out. Thus we have shown that $\partial_s G = 0$ for all $s$. Since $G(0) = P_0$, we see that
\begin{align}
    G(s) = P_0, \  \forall\, s,
\end{align}
confirming that our definition satisfies eqn.~\ref{eqn:U_A_def}.\par
Now all that remains is to show that, as $ T \rightarrow \infty$, the actual dynamics, $U_T$, asymptotically approach the adiabatic dynamics described by $U_A$. We will omit the proof of this result, which is simply a proof of the adiabatic theorem, however a complete justification can be found in~\cite{avron_adiabatic_1987}. We quote the result, which compares the projector generated by non-adiabatic dynamics, $P(s) = U_T P_0 U^{\dag}_T$ compared to that generated by the adiabatic time evolution $P_I(s) = U_A P_0 U^{\dag}_A$. The result states that
\begin{align}
    |P(s) - P_I(s)| = \mathcal O (T^{-1})
\end{align}
Thus we can see that in the limit of large $T$, $U_T$ generates the correct time evolution.

\section{Adiabatic Current Operators}\label{sec:near_adiabatic}

In this section we will expand our justification for the form of the adiabatic current presented in \textsection \ref{sec:preliminaries}. The argument splits into two parts. First we introduce an operator that describes the direct flow of particles between a pair of individual sites $\textbf r_i$ and $\textbf r_{j}$ in the system, $f_{\textbf r_i , \textbf r_j}$. Then we will show how, in the adiabatic limit, the flow between a pair of sites is determined by the adiabatic Hamiltonian $K$, introduced in \textsection\ref{sec:kato}. Bulk current can always be written as the sum over the flow between individual sites, thus all the properties of $f_{\textbf r_i , \textbf r_j}$ naturally extend to the currents defined in~\textsection\ref{sec:current}. 

\subsection{Two-Point Flow Operator}

The flow between individual sites can be extracted from the continuity equation for electron density. We start by looking at the change in occupation number at a single site, $\delta_{\textbf r} = \ket {\textbf r} \bra{\textbf r}$, for a projector onto a set of occupied states $P$,
\begin{align}
    \partial_t \left \langle \delta_{\textbf r} \right \rangle = \Tr \left (
    \partial_t P \delta_{\textbf r} 
    \right ).
\end{align}
We use the Von-Neumann equation to express $\partial_t P$ in terms of the Hamiltonian, cycling the terms in the trace to arrive at
\begin{align}
    \partial_t \left \langle \delta_{\textbf r }\right \rangle = \Tr \left (P i [H,\delta_{\textbf r}]\right ).
\end{align}
Inserting a complete set of position eigenstates $\1 = \sum_{j} \delta_{\textbf r_j}$, we may re-express this in the form of a continuity equation,
\begin{align}\label{eqn:continuity_eqn}
    \partial_t \left \langle \delta_{\textbf r }\right \rangle = 
    \sum_{j} \Tr \left [ P (
    i \delta_{\textbf r_j}H\delta_{ \textbf r }
    -i\delta_{\textbf r}H\delta_{ \textbf r_j }
    )
    \right ].
\end{align}
This describes the change of number density at site $\textbf r$ as a sum of the flow from every other site $\textbf r_j$ in the system. Thus we can interpret the operator 
\begin{align}\label{eqn:two_point_current}
    f_{\textbf r_i,\textbf r_j} = i \left (\delta_{\textbf r_j}H\delta_{\textbf r_i }-\delta_{\textbf r_i}H\delta_{ \textbf r_j }\right ),
\end{align}
as quantifying the flow of electrons from the site at $\textbf r_i$ to $\textbf r_j$. Note that it is antisymmetric in $\textbf r_i \leftrightarrow \textbf r_j$.\par
In the next section we will work in the adiabatic limit, thus let us find the current per unit of scaled time $s$. Using the substitution $t = sT$, we define the scaled flow operator
\begin{align}
    \bar f_{\textbf r_i,\textbf r_j} = iT \left (\delta_{\textbf r_j}H\delta_{\textbf r_i }-\delta_{\textbf r_i}H\delta_{ \textbf r_j }\right ),
\end{align}
which satisfies the scaled continuity equation $\partial_s \left \langle \delta_{\textbf r} \right \rangle = \sum_j \Tr (P \bar f_{\textbf r_i,\textbf r_j})$.\par
Both current operators described in \textsection\ref{sec:current} can be written as a sum of individual flows. The operator in def.~\ref{eqn:non_resolved_current}, representing the total current between the two halves of the system, is the sum of all two-point currents where $\textbf r_i \rightarrow \textbf r_j$ crosses $x = R_x$,
\begin{align}
    J_{R_x} &= \sum_{\substack{x_i > R_x \\ x_j < R_x} } f_{\textbf r_i,\textbf r_j},
\end{align}
whereas def.~\ref{eqn:current_operator} may be written as the sum of all two-point currents that cross $x = R_x$ and either start or finish at $\textbf r$,
\begin{align}
    J_{R_x}(\textbf r) = \sum_{x_i > R_x} f_{\textbf r, \textbf r_i},
\end{align}
where we have assumed that $\textbf r$ has $x<R_x$.

\subsection{Adiabatic Current}

Here we show that the scaled flow operator may be replaced with the adiabatic equivalent up to small corrections in $T^{-1}$:
\begin{align}
    \bar f_{\textbf r_i, \textbf r_j} = \bar f_{\textbf r_i,\textbf r_j}^A + \mathcal{O} \left ( \frac{1}{T} \right),
\end{align}
with $f^A$ defined according to
\begin{align}\label{eqn:adiabatic_current}
        \bar f_{\textbf r_i,\textbf r_j}^A = i \delta_{\textbf r_j} \bar K\delta_{\textbf r_i }-i\delta_{\textbf r_i}\bar K\delta_{ \textbf r_j },
\end{align}
and $\bar K = i[\partial_s P_I , P_I] $ in analogy to the definition in \textsection \ref{sec:kato}. Note that the flow in eqn.~\ref{eqn:two_point_current} is always valid regardless of what state it is acted on. Since $K$ is defined for a choice of $P_I$, the adiabatic flow is only correct when evaluated with an occupied band consisting of the states in $P_I$.\par
Our argument has two parts. We shall first decompose the Hamiltonian into a term that commutes with $P$ and a term that does not. Only the non-commuting term contributes to measurable change in the system. Finally, we show that this term has a straightforward expression in the near-adiabatic limit, and defines the adiabatic flow. \par
As shown in appendix \ref{sec:kato_appendix}, the time-evolution of our system is determined by the scaled Von-Neumann equation,
\begin{align}\label{eqn:von_neumann_ap}
    \partial_{s} P = -i T[H,P],
\end{align}
where we treat $T$ as a tunable parameter. Clearly, any component in $H$ that commutes with $P$ will vanish in this expression. Thus, let us split $H$ into a commuting and non-commuting part,
\begin{align}
    H = (PHP + QHQ) + (PHQ + QHP),
\end{align}
where we define $Q = \1 - P$ as the projector onto unoccupied states. We will label the two terms in brackets $H_{\parallel}$ and $H_{\perp}$ respectively. $H_{\parallel}$ commutes with $P$, vanishing in the Von-Neumann equation and thus does not effect any change in $P$ whatsoever. Furthermore, any Hamiltonian that differs from H only by a term that commutes with $P$ (e.g. $H' = H + PMP + QMQ$ for some arbitrary operator $M$) will still produce the exact same dynamics for $P$.\par
Using these two components of $H$, we can split the flow operator into two components according to
\begin{align}
\begin{aligned}
    \bar f_{\textbf r_i, \textbf r_j} = 
    &\ (iT \delta_{\textbf r_j}H_\parallel \delta_{\textbf r_i } + h.c.)\\ 
    &+ 
    (iT \delta_{\textbf r_j}H_\perp \delta_{\textbf r_i } + h.c.),
\end{aligned}
\end{align}
and label these two terms as $\bar f^\parallel _{\textbf r_i, \textbf r_j}$ and $\bar f^\perp_{\textbf r_i, \textbf r_j}$ respectively. Let us examine the contribution of each term to the continuity equation \ref{eqn:continuity_eqn},
\begin{align}
    \partial_s \langle \delta_{\textbf r}\rangle = \sum_j \Tr \left [ 
    P \left (\bar f^{\parallel}_{\textbf r, \textbf r_j} + 
     \bar f^{\perp}_{\textbf r, \textbf r_j}\right )
    \right ].
\end{align}
Writing out the $\bar f^\parallel$ term explicitly, we see that it vanishes
\begin{align}\label{eqn:f_vanishes}
    \begin{aligned}
    \sum_j \Tr \left [ 
    P \bar f^{\parallel}_{\textbf r, \textbf r_j}
    \right ] &  = 
    \sum_j \Tr \left ( iT P 
    \delta_{\textbf r_j} H_{\parallel}\delta_{\textbf r}
    \right ) + h.c. \\ 
    & = \Tr \left ( 
      iT PHP \delta_{\textbf r}
    \right ) + h.c. \\ 
    &= 0.
    \end{aligned}
\end{align}
Thus, the only term that contributes to the continuity equation is $\bar f^\perp_{\textbf r_i, \textbf r_j}$. We can interpret $\bar f^\parallel_{\textbf r_i, \textbf r_j}$ as a circulating current, representing the flow between occupied states in $P$. As it vanishes in the continuity equation, it is not possible for this term to change the number density at any site in the system. \par
Now, let us examine $\bar f^\perp_{\textbf r_i, \textbf r_j}$ in the near-adiabatic limit. As we have seen in appendix \ref{sec:kato_appendix}, the projector is equal to the instantaneous projector $P_I$ up to order $T^{-1}$. Thus, treating $T$ as a tunable parameter, let us expand the dynamics around the limit $T \rightarrow \infty$, 
\begin{align}\label{eqn:p_expansion}
    P = P_I + \frac{1}{T}\delta P + \mathcal{O}\left (\frac{1}{T^2}\right ),
\end{align}
where $\delta P$ is the first-order perturbation away from the exact adiabatic result. Inserting this into eqn~\ref{eqn:von_neumann_ap}, we get
\begin{align}
    \partial_s P_I + \frac{1}{T} \partial_{s}\delta P = -iT [H, P_I] -i [H, \delta P] + ...
\end{align}
The terms omitted are at least of order $T^{-1}$. The first term on the right vanishes, since $P_I$ and $H$ commute by definition. Thus, collecting the next lowest order in $T$, we get
\begin{align}
    \partial_s P_I = -i [H, \delta P].
\end{align}
Time evolution of $P_I$ is determined by $\bar K$, thus we arrive at the identity
\begin{align} \label{eqn:HPKP}
    [H, \delta P] = [\bar K, P_I].
\end{align}\par
Let us consider $H_\perp$ in this limit. It is straightforward to show that
\begin{align}
\begin{aligned}
    H_{\perp} &= PHQ+QHP \\ 
    &= \big[ [H,P] , P  \big].
\end{aligned}
\end{align}
$H$ commutes with $P_I$, so we can set  
\begin{align}
    [H,P] =[H, \frac{1}{T} \delta P]+ 
    \mathcal{O} \left (\frac{1}{T^2} \right ),
\end{align}
and use identity \ref{eqn:HPKP} to get this in the form.
\begin{align}
    H_{\perp} &= \frac{1}{T}  \big[ [\bar K, P_I] , P_I  \big] + 
    \mathcal{O} \left (\frac{1}{T^2} \right ), 
\end{align}
where we have also expanded the second $P$ according to eqn.~\ref{eqn:p_expansion}. Finally we use the identities $\partial_s P_I = -i [\bar K, P_I]$ and $\bar  K = i[\partial_s P_I, P_I]$ to show that
\begin{align}
    H_{\perp} = \frac{1}{T} \bar K + \mathcal{O} \left (\frac{1}{T^2} \right).
\end{align}
Thus, in the near-adiabatic regime, we have shown that the flow operator separates into two terms
\begin{align}
    \bar f_{\textbf r_i,\textbf r_j} = \bar f^\parallel_{\textbf r_i,\textbf r_j} + \bar f^\perp_{\textbf r_i,\textbf r_j}
\end{align}
The first represents a circulating current that does not effect a change in $P$ whatsoever. In any sum of current into or out of a region, we expect any $f^\parallel$ terms to vanish, since the total flow from this term into any individual site vanishes according to eqn.~\ref{eqn:f_vanishes}. Additionally, it is worth noting that the term,
\begin{align}
    \bar f^\parallel _{\textbf r_i, \textbf r_j} = iT \delta_{\textbf r_j}H_\parallel \delta_{\textbf r_i } + h.c.
\end{align}
is linearly dependent on $T$, since $H^\parallel$ has nonzero components of order $T^0$. Thus it diverges as $T \rightarrow \infty$. Any constant current scaled over an infinitely large time period will diverge. In what follows, we shall omit these terms, meaning that our current operator will not account for persistent or \textit{frozen in} current in the bulk at any point in the adiabatic evolution.\par
The second term, given by
\begin{align}
    \bar f^\perp_{\textbf r_i, \textbf r_j} = i \delta_{\textbf r_j}\bar K\delta_{\textbf r_i }-i\delta_{\textbf r_i}\bar K\delta_{ \textbf r_j },
\end{align}
represents the current generated over the course of the adiabatic evolution by the change in the electronic arrangements of states in the band, $P$. Thus, we see that in the adiabatic limit, in the absence of persistent current, the flow in the system can be approximated by the adiabatic flow operator,
\begin{align}
    \bar f_{\textbf r_i, \textbf r_j} = i \delta_{\textbf r_j}\bar K\delta_{\textbf r_i }-i\delta_{\textbf r_i}\bar K\delta_{ \textbf r_j } + \mathcal O \left( \frac{1}{T} \right ).
\end{align}
Finally, we may rescale from $s$ back to $t$, making the substitutions $\bar f = f T$ and $\bar K = K T$ to arrive at
\begin{align}
    f_{\textbf r_i, \textbf r_j} = i \delta_{\textbf r_j} K\delta_{\textbf r_i }-i\delta_{\textbf r_i} K\delta_{ \textbf r_j } + \mathcal O \left( \frac{1}{T^2} \right ).
\end{align}
Since the current operators used in \textsection \ref{sec:preliminaries} and \textsection\ref{sec:marker_derivation} can be expressed as a sum over individual flows, this property naturally extends to the current, and we see that
\begin{align}
    J_{R_x} &\rightarrow J^A_{R_x} + \mathcal O \left( \frac{1}{T^2} \right ) \\
    J_{R_x}(\textbf r) &\rightarrow J^A_{R_x}(\textbf r) + \mathcal O \left( \frac{1}{T^2} \right ),
\end{align}
where in each case, the adiabatic version is found by substituting $H \rightarrow K$ in the definition.

\section{Connection to Chern Marker} \label{sec:chern_marker_connection}

The crosshair marker can be seen as an extension of the Chern marker derived in~\cite{bianco_mapping_2011}. To show this let us look at what happens when we integrate a step function over an interval $[0,L]$.
\begin{align}
    \int_0 ^{L_x} \theta(x - R_x) d R_x =  x \text{ for } 0< x < L_x.
\end{align}
Note that this is an integral even in a lattice system, since the parameter $R_{x}$, which sets the position of the step function, is always continuous. Thus let us assume that our quantum system is on a rectangular region of size $L_x, L_y$. We wish to integrate the cross-hair marker in $\textbf p$ over this region, essentially summing the marker evaluated about every point,
\begin{align}
    \int_{\textbf L} \sigma(\textbf r, \textbf R) d^2 \textbf R & = 2 \Im \Tr_{\textbf r}\int d^2\textbf R   P\theta_{R_x} P \theta_{R_y} P \\
    & =  2\Im \Tr_{\textbf r}  \left ( P X P Y P \right ).
\end{align}
Thus, we arrive at the standard form of the Chern marker. The cross-hair marker can be understood as a deconstruction of the Chern marker that allows us to extract a quantised result for the Chern number of a non-uniform system.

\begin{figure}[ht!]
\centering
\includegraphics[width = \columnwidth]{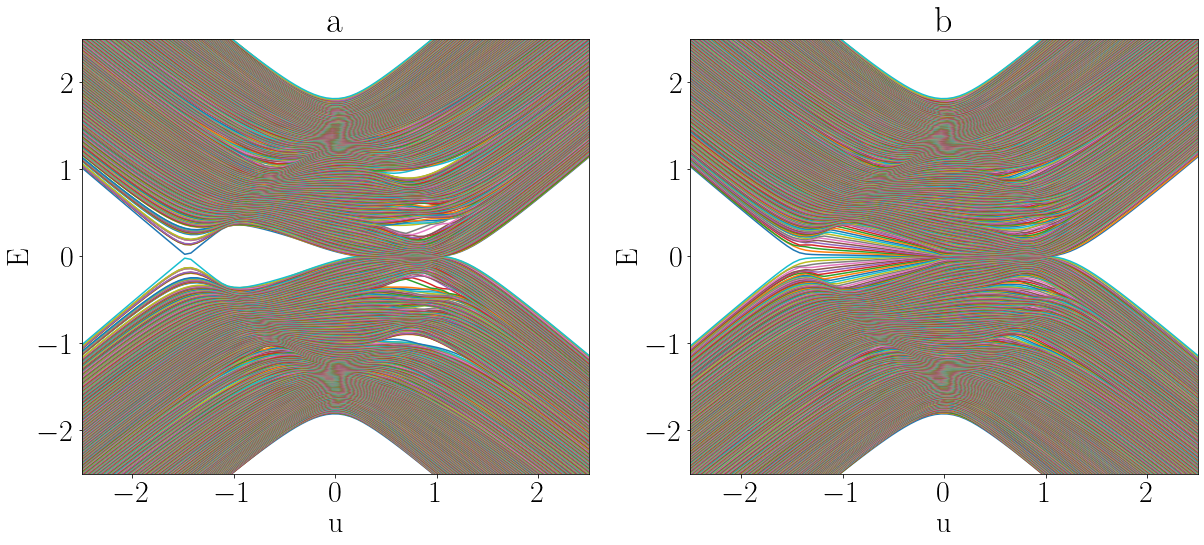}
\caption{\label{fig:band_graphs} Energy of the states of an amorphous QWZ system with 800 sites on a random Voronoi lattice plotted against the internal parameter $u$. a) The system in periodic boundaries. b) The system in open boundaries. }
\end{figure}

\section{Edge States in Amorphous Hamiltonians}\label{sec:amorphous_appendix}

\begin{figure*}
\centering
\includegraphics[width = 0.75\paperwidth]{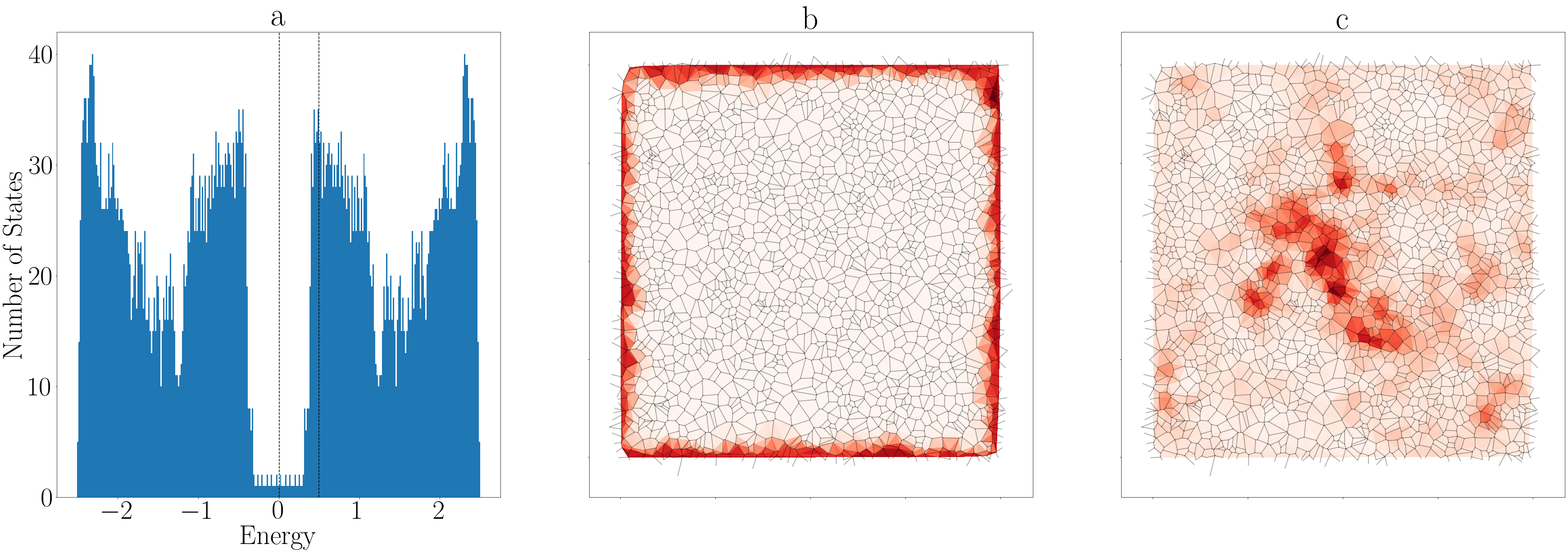}
\caption{\label{fig:states_examples} a) Density of states for a $40\times40$ amorphous QWZ system in open boundaries with $u = -1$. The two vertical dashed lines indicate the energy of the states shown in subplots b and c. b) Heatmap of the probability amplitude, $|\psi(\textbf r)|$, for an edge state with energy $-0.006$. c) Heatmap of the probability amplitude for a bulk state with energy $0.5$. In plots b and c we have superimposed the lattice structure in black. }
\end{figure*}

The topological phase diagram of the amorphous QWZ model is verified by looking at the band structure in periodic and open boundary conditions as a function of the parameter $u$. If the system is a topologically trivial insulator, we expect that it should be gapped for periodic boundary conditions and that the gap should remain open when we move to open boundaries. On the other hand, if the system is a topological insulator then the gap will be closed in open boundaries by the presence of edge modes. Furthermore we should always expect the gap to close whenever $u$ crosses between a topological and trivial phase. The energies as a function of $u$ are shown in figure \ref{fig:band_graphs}. As can be seen, a topological gap is open for $-1.5 < u < 0$ at zero Fermi level. In figure~\ref{fig:states_examples}, we show the density of states for a system with $u = -1$ alongside the probability amplitudes $|\psi(\textbf r)|$ for two example states, one edge state and one bulk state.

\end{document}